# On the temperature dependence of the hopping thermopower in disordered semiconductors

O.E. Parfenov and F.A. Shklyaruk

*Russian Research Center Kurchatov Institute, 123182, Moscow, Russia*

(Dated: October 23, 2006)

At low temperature using thermodynamics of irreversible processes the general expressions for the temperature dependence of the thermopower in the case of the hopping conductivity for disordered materials are found. The account of influence of impurity levels degeneration on the thermopower is lead. In a view of received results experimental data of the thermopower in amorphous and impurity semiconductors are discussed.

**1. Introduction.**

Typical feature of disordered semiconductors, to which concern amorphous and impurity semiconductors, is presence of localized states (the local centers), located in the band gap. At low temperature transport of a charge is carried out by hopping between localized states. The probability of a jump is defined by probability to absorb phonon with the energy $\varepsilon$

$$\exp(-\varepsilon/kT) \qquad (1)$$

and the factor of wave functions overlapping

$$\exp(-2r/a_L), \qquad (2)$$

where $r$ is distance between localized states, and $a_L$ is localization radius. The competition between these two factors results to two possible types of the hopping conductivity. If $\varepsilon/kT < 2r/a_L$ when jumps occur between neighbors states, if $\varepsilon/kT > 2r/a_L$ when the variable-range hopping conductivity takes place.

Mott has shown [1] that not all of states give contribution to conduction, but only states with energy in active band $2\Delta$ near Fermi level, where $\Delta$ is characteristic energy of a jump. In three-dimensional case he derived for the variable-range hopping conductivity

$$\sigma = \sigma_0 \exp\left(-(T_0/T)^{1/4}\right), \qquad (3)$$

where $\sigma_0$ is the pre exponential factor, which weakly depends on temperature. Energy of a jump $\Delta$ is equal $k(T_0 T^3)^{1/4}$. The constant $T_0$ is directly connected to density of states at the Fermi level and radius of localization:

$$T_0 = \frac{\gamma}{kN(\mu)a_L^3}, \qquad (4)$$

where $N(\mu)$ is density of state at Fermi level and $\gamma = Const$.

Owing to electron-electron interactions form of the columb gap in density of states at Fermi level is possible. It is lead to different from (3) dependences of conductivity [2]

$$\sigma \propto \exp\left(-(T_0/T)^{1/2}\right). \qquad (5)$$

In the case of the nearest neighbors hopping, conductivity has activation form with activation energy $\varepsilon_3$

$$\sigma = \sigma_3 \exp(-\varepsilon_3/kT). \qquad (6)$$

To receive exact value of factor $\gamma$ in equation (3) or dependence of hopping conductivity from impurity quantity and form of the density of states the percolation theory methods where used [3]. Employments of percolation theory methods get possible after work of Miller and Abrahams [4], they show that solution of the problem of hopping conductivity in disordered materials can be reduce to calculation conductivity of a random network of resistance. Each unit of the network corresponds to one local center, and units are in pairs connected resistance.

For a finding of the temperature dependence of the thermopower the percolation theory for the first time has been used by Zvyagin [5]. In case of the variable range hopping conductivity and if a density of states slowly varies within the active band, the thermopower is equal

$$\alpha = -\frac{1}{e}\xi \frac{(T_0 T)^{1/2}}{\widetilde{E}}, \qquad (7)$$

where $\xi$ is constant of the order of one tenth, $\widetilde{E} = (d\ln N(E)/dE)^{-1}_{E=\mu}$ is characteristic energy at which the density of states varies. We shall note, if density of states is a constant the thermopower become zero. In case of the nearest neighbors hopping conductivity thermopower is equal [6]

$$\alpha = -\frac{1}{e}\frac{\varepsilon_3}{T} + A, \qquad (8)$$

where $\varepsilon_3$ is activation energy, $A$ is kinetic member which doesn't depend on temperature.

Different from (8) result is obtained [7] by means of the effective medium method, in which resistances between neighbor states average by all pairs

$$\alpha = -\frac{5}{6}\left(\frac{E_m^2(0)}{2kT} + \frac{2\pi kT}{3}\right)\left(\frac{d\ln N(E)}{dE}\right)_{E=\mu}, \quad (9)$$

where $E_m(0)$ is activation energy of conductivity at zero temperature. However the author does not give clear substantiation of the linear temperature dependence of thermopower received him, that does not allow to use the formula (9) with confidence.

The analysis of experimental data shows, that temperature dependences of thermopower a kind (7,8) in exact form are observed seldom. In case of variable-range hopping conductivity the thermopower has form like $A\sqrt{T} + BT$ [8-12], and in case the nearest neighbors hopping conductivity $A/T + BT$ [13-15]. Cleanly linear temperature dependence [16,17] is often observed also. Similar behavior is peculiar to metals, in which thermopower is described by Mott formula [1]

$$\alpha = \frac{\pi^2}{3}\frac{k}{e}\frac{kT}{\widetilde{E}}. \qquad (11)$$

This formula was derived for delocalized states and is not applicable in case of the hopping conductivity. Besides the described cases, at low temperature temperature independent thermopower is observed [18,19].

Thus, the percolation theory, well describing temperature dependence of hopping conductivity [1-3], only partially describes thermopower.

The purpose of this work is the explanation observable in experiments temperature dependences of the thermopower in case of the hopping conductivity, in particular an explanation of the "metal" dependence of the thermopower.

The phenomenon of the hopping conductivity at direct current we consider as the process of the macroscopical electron diffusion. It allows to use methods of irreversible processes thermodynamics and statistical physics to calculate thermopower. Such approach allows don't work out details of impurity state type. We considered only diffusion thermopower, as for hopping mechanism charge transfer the effect of phonon drag is absent [5].

## 2. Hopping conductivity as macroscopical diffusion.

In the continuous medium diffusion is described by random walker and an average square of displacement is proportional to time

$$<r^2> \propto t. \qquad (11)$$

While to the percolation theory corresponds the anomaly law of diffusion with slower time dependence [20]

$$<r^2> \propto t^k, k < 1, \qquad (12)$$

that caused fractal structure of the percolation cluster. In case of classical diffusion probability of transition remains to the same after each jump, irrespective of position of a particle. In case of percolation each jump is characterized by individual probability of a jump and depends from previous transitions.

The percolation theory consider a medium, in which bound between two sites is whole randomly with probability $p$, and broken off with probability $(1-p)$. Then if $p = 0$ all bounds are broken off, and if $p = 1$ medium consists of whole bounds. There is some critical value $p_c$ called the percolation threshold, at which for the first time the opposite edges of the considered medium connected by whole bounds [2].

At calculation of conductivity with percolation theory consider, that the system is close percolation transition. As beginning from the moment when $p = p_C + \delta$, where $\delta$ define critical region, by removing from threshold conductivity almost doesn't change [2]. But real samples can be in the distance from the percolation threshold. In this case a crossover form percolation process to normal diffusion can have place, when scale of considered systems much more exceeds of the correlation length of percolation cluster $\xi_c$.

Let's define, when the normal diffusion will be carried out. The sample with line scale $l \sim 1\,\text{cm}$ is possible to break for many $\sim 10^{12}$ macroscopical parts with size $h \sim 1\,\mu\text{m}$, in which thermodynamical fluctuations is small [21]. The size of the critical area for an elementary micrometer volume is defined by $\delta = (a/h)^{1/\nu}$ [2], where $a$ is a distance between impurity sites and $\nu$ is a critical index (equal 0.88 in three-dimensional case). If $a$ is less than $h/100$, we receive $\delta \leq 10^{-2}$. In case of the nearest neighbor jumps, for which the correlation length is equal $\xi_c = a(2r_c/a + \varepsilon_3/kT) \sim N^{-1/3}$ and $p - p_C = (a/\xi_c)^{1/\nu} \approx 0.1 > \delta$, the condition of crossover is $N > 10^{12}\,\text{см}^{-3}$. In case of the variable range jumps $\xi_c = a(T_0/T)^{(1+\nu)/4}$, for $p - p_C = (a/\xi_c)^{1/\nu} \approx 0.1 > \delta$ the condition of crossover is $T_0/T < 10^6$. We can expect, outside of the critical region the number of jumps, for which a charge passes all sample under electric field, become very large. At $h \gg \xi_c$ with increasing the number of jumps of a charge occurs averaging of jump probability on all parameters of transport of a charge. So, in a large scale the complex in microscopical structure percolation process turns into normal.

Thus, at received conditions the hopping conductivity can be considered as process of normal diffusion on large distances. It enables to use the irreversible thermodynamics when we consider thermopower in case of the hopping conductivity.

### 3. Thermopower in case of the hopping conductivity.

The thermodynamics of irreversible processes is based on concepts of local balance and Onsager relations [21]. From a principle of symmetry of the kinetic coefficient formulated in general view by Onsager, Calvin's relation between thermopower and Peltie's coefficient $\Pi$ follows

$$\alpha = -\frac{\Pi}{eT}. \quad (13)$$

In turn, Peltie's factor is defined by average energy, transported by charge carriers and measured from the chemical potential $\mu$. Substituting $\Pi = <E - \mu>$ in (13) we shall receive

$$\alpha = -\frac{k}{e}\frac{<E - \mu>}{kT}. \quad (14)$$

Energy averaged by a current of particles is

$$<E - \mu> = \frac{\int\limits_0^{+\infty}(E - \mu)dj_n(E)}{\int\limits_0^{+\infty}dj_n(E)}. \quad (15)$$

The current density of charge carriers $\int\limits_0^{+\infty}dj_n(E)$ is defined by drift velocity $V_{dr}$, density of states $N(E)$, Fermi distribution $f(E)$ and probability $p(E)$ that the particle with energy $E$ participates in hops. Thus the current density of particles with energy from $E$ to $E + dE$ is

$$dj_n(E) = -eV_{dr}N(E)f(E)p(E)dE. \quad (16)$$

For normal diffusion from the probability theory follows, that $p(E)$ looks like Gauss function. We shall choose so that average energy of a jump it was equal $\Delta$

$$\int\limits_0^{+\infty}E^2 p(E)dE = \Delta^2. \quad (17)$$

Let's rewrite the formula (15) substituting $dj_n(E)$ and $p(E)$ in it

$$<E - \mu> = \left[\int\limits_0^{+\infty}(E - \mu)N(E)f(E)\right.$$
$$\left.\times \exp\left(-\frac{(E - \mu)^2}{2\Delta^2}\right)dE\right]$$
$$\div \left[\int\limits_0^{+\infty}N(E)f(E)\exp\left(-\frac{(E - \mu)^2}{2\Delta^2}\right)dE\right]. \quad (18)$$

While we have considered only energy transferable by electrons, but it is necessary to consider holes also. Electrons and holes contributions to thermopower in case of the hopping conductivity don't additive [22]. Then average transferable energy is

$$<E-\mu> = \frac{1}{2}\left(<E-\mu>_n + <E-\mu>_p\right). \quad (19)$$

At low temperatures $(kT << \mu)$ it is possible to be limited to the first temperature member in expansion of integral

$$\int_0^{+\infty} G_i(E)f(E)dE \approx \int_0^\mu G_i(E)dE + \frac{\pi^2}{6}(kT)^2\left(\frac{dG_i(E)}{dE}\right)_{E=\mu}, \quad (20)$$

where $f(E)$ is Fermi distribution function, and function $G_i(E)$ in our case is equal $(E-\mu)N(E)\exp[-(E-\mu)^2/2\Delta^2]$ and $N(E)\exp[-(E-\mu)^2/2\Delta^2]$ for numerator and denominator in (18) accordingly.

Including, that the density of state slowly varies near to the Fermi level, we shall receive

$$<E-\mu> \approx \left(\frac{\pi-2}{\pi}\Delta^2 + \frac{2\pi}{3}(kT)^2\right) \times \left(\frac{d\ln N(E)}{dE}\right)_{E=\mu}. \quad (21)$$

Substituting (21) in (14) we find, that hopping thermopower is

$$\alpha \approx -\frac{k}{e}\left(\frac{\pi-2}{\pi}\frac{\Delta^2}{kT} + \frac{2\pi}{3}kT\right) \times \left(\frac{d\ln N(E)}{dE}\right)_{E=\mu}. \quad (22)$$

In three-dimensional case, if the variable range hopping conductivity takes place, the characteristic transferable energy $\Delta$ is equal $k(T^3T_0)^{1/4}$. Substituting $\Delta$ in (22) we shall receive

$$\alpha \approx -\frac{k}{e}\left(\frac{\pi-2}{\pi}k\sqrt{TT_0} + \frac{2\pi}{3}kT\right) \times \left(\frac{d\ln N(E)}{dE}\right)_{E=\mu}. \quad (23)$$

In case of two-dimensional conductivity when $\sigma \propto \exp(-T^{-1/3})$ and $\Delta = k(T_0T^2)^{1/3}$ thermopower is

$$\alpha \approx -\frac{k}{e}\left(\frac{\pi-2}{\pi}k\sqrt[3]{TT_0^2} + \frac{2\pi}{3}kT\right) \times \left(\frac{d\ln N(E)}{dE}\right)_{E=\mu}. \quad (24)$$

Also formation of the coulomb gap in density of states at the Fermi level is possible, in this case $\Delta = k(T_0T)^{1/2}$ and conductivity has form (5). Considering, that near to the Fermi level $N(E) \propto E^2$ and recalculating (19), we receive zero thermopower. It is consequence of symmetry of transitions electrons and holes concerning to the Fermi level.

If in the case of the nearest neighbor hopping conductivity condition $kT << \mu$ is satisfied, that, substituting $\Delta = \varepsilon_3$ in (22), we shall receive

$$\alpha \approx -\frac{k}{e}\left(\frac{\pi-2}{\pi}\frac{\varepsilon_3^2}{kT} + \frac{2\pi}{3}kT\right) \times \left(\frac{d\ln N(E)}{dE}\right)_{E=\mu}. \quad (25)$$

It is visible, that the thermopower has a minimum at

$$T_{min} = \varepsilon_3\sqrt{\frac{3(\pi-2)}{2\pi^2}} \quad (26)$$

and if we define temperature of a minimum, it is possible to estimate activation energy of conductivity.

**4. The thermopower caused by degeneration.**

Till now we did not consider degeneration of the local centers, which can appear essential for the thermopower. Some authors [1,5] specify necessity of the account of the contribution in hopping thermopower related to degeneration of impurity levels. So according to [1] for non-magnetic materials at high temperatures the entropy contribution equal to $(k/e)\ln 2 \approx 60\ \mu V/K$ should be present. For small polarons this question was considered in work [23] for magnetic semiconductors where it is shown, that the contribution connected to degeneration is not displayed almost because of spin-polaron effects both in paramagnetic region and in the region with magnetic order.

The Fermi distribution in view of degeneration

$$f(E) = \frac{1}{1 + \beta e^{-(E-\mu)/kT}}, \quad (27)$$

where $\beta$ is the factor of degeneration of localized states including both spin and orbital contributions [24]. Let's redefine chemical potential

$$\widetilde{\mu} = \mu - kT \ln \beta. \quad (28)$$

As thermopower is

$$\alpha = -\frac{k}{e}\frac{<E-\mu>}{kT},$$

in view of (28) we receive the additional thermopower caused by degeneration

$$\alpha_{deg} = -\frac{k}{e}\ln \beta. \quad (29)$$

Degeneration should also influence on the average energy $<E>$ in consequence of change the chemical potential. The account of degeneration results to additional thermopower

$$\Delta\alpha_{deg} \approx -\frac{k}{e}\ln \beta$$
$$\times \left(\frac{\pi-2}{\pi}\Delta^2 + \frac{2\pi}{3}kT^2\right)\widetilde{E}^{-2}. \quad (30)$$

However these corrections are negligible. If to take for estimation $T_0 = 10^5$ K, $\widetilde{E} = 100$ meV for the variable range hopping conductivity at 10 K, ration $\Delta\alpha_{deg}/\alpha_{deg}$ is equal 0.003. Influence $\Delta\alpha_{deg}$ on temperature dependence of thermopower is insignificant too. So at specified above conditions the temperature addition is less than 1%. Thus, it is possible to distinguish the contribution $\alpha_{deg}$ and used (29) to define the degeneration factor $\beta$. Displays full contribution $\alpha_{deg}$ is possible only under condition when the local center is paramagnetic.

## 5. Discussion.

From received above formulas (23-25) follows, that thermopower adds from two contributions. The first is defined by carry of potential energy of carriers, and the second is related with thermal energy. Last circumstance is not considered by the percolation theory for the thermopower.

The quantity of constant $T_0$ defines prevalence either root or linear member in temperature dependences of thermopower. So for $T < 100$ K at $T_0 > 10^4$ K dominates root temperature dependence, and at $T_0 < 10^3$ K thermopower has almost linear temperature dependence. The received result will qualitatively be corresponded with experimental data. In work [8] thermopower of a-GaSb was investigated. At temperature below 70 K conductivity follows the Mott law (3) with $T_0 = 7 \cdot 10^4$ K. The thermopower as well as follows from (23) has root temperature dependence. Influence of quantity of an impurity for the thermopower can be traced by work [12]. This work contains the review of researches of thermopower in $CH(FeCl_2)_x$ with different dopping levels. The author has empirically shown, that experimental curves of temperature dependence of thermopower look like $A\sqrt{T} + BT$, that coincides with received us formula (23). From formula (23) follows that the increase quantity of an impurity bring about increasing roles of the linear member and to reduction of thermopower value, that correspond with experiment. At increasing $T_0$ temperature dependence of thermopower in formula (23) comes near to almost root. As shown above, our approach is true only when $T_0/T < 10^6$. When $T_0/T > 10^6$ correctly to use the formula (7), so as this region corresponds to percolation process of carry of a charge with the very large correlation length.

In case of the nearest neighbor hopping conductivity thermopower is described by the formula (25) and has the minimum (26), that is often observed in experiment. In review [14] transport properties of $(Pb_{0.78}Sn_{0.22})_{1-y}In_yTe$ are resulted. For samples with $y = 2\%$ and 3% thermopower a kind AT+B/T was observed. From conductivity data value of activation energy $\varepsilon_3$ was found. For sample with $y = 3\%$ $\varepsilon_3 = 44\,\text{meV}$. The minimum of thermopower corresponds approximately to 200 K, from (26) we found $\varepsilon_3 \approx 44\,\text{meV}$. In work [13] $Fe_3O_{4-x}F_x$ were investigated with different dopping levels, for all samples the minimum of thermopower in the nearest neighbor hopping conductivity region were found. For sample with $x = 0.1$ from conductivity data we have $\varepsilon_3 = 61\,\text{meV}$ and from the minimum of thermopower (26) we find $\varepsilon_3 \approx 60\,\text{meV}$. For sample with $x = 0.05$ we find $\varepsilon_3 = 76\,\text{meV}$ and $\varepsilon_3 \approx 40\,\text{meV}$ from conductivity and from minimum of thermopower correspondingly.

At low temperature in neutron-alloyed Ge:Ga [25] and in quasiamorphous carbons [26] jump of thermopower to zero is observed. At same time, conductivity is carried out by variable range jumps according (5). As it has been shown above, in case of presence symmetric "coulomb gap" at the Fermi level, thermopower should equal to zero.

The thermopower of amorphous germanium [18] in the field of 60-300K appears not to be dependent on temperature and is approximately equal to $60\,\mu\text{V/K}$. Independent on temperature thermopower can be connected with the constant density of states near to the Fermi level. Then the thermopower in conformity with the formula (22) become equal to zero and there is only a member caused by degeneration (29) equal $k/e \cdot \ln 2 \approx 60\,\mu\text{V/K}$, which is close to experimental values of the thermopower. Researches of $La_2CuO_4$ [27,28] show anomaly behavior of thermopower in magnetic transition region. Depending on oxygen concentration, jumps of thermopower in antiferromagnetic transition region $60\,\mu\text{V/K} = k/e \cdot \ln 2$ or $120\,\mu\text{V/K} = k/e \cdot \ln 4$ are observed, or above $T_N$ the wide plateau about $\sim 120\,\mu\text{V/K}$ is observed.

## 6. Conclusion.

Within the range of irreversible thermo-dynamic at low temperatures hopping thermopower in disordered semiconductors is calculated. It has allowed to explain linear temperature dependence of thermopower observable in experiments. The reason of its occurrence is asymmetry transferable by electrons and holes energy as owing to changeable density of states close to the Fermi level. In case of small concentration and at low temperatures root temperature dependence of thermopower is played defining role, thus normal diffusion on large scales should change to percolation. At weak interaction of impurity state spin with an environment it is possible to receive data about degeneration of impurity levels.

Comparison with experiment has shown, that in many cases received above formulas qualitatively describe experimental data.

Authors are grateful I.P. Sadikov and A.A. Chernishov for attention and support of this work.